# Cloaking the Magnons

Mehrdad Elyasi, Charanjit S. Bhatia, Cheng-Wei Qiu, and Hyunsoo Yang[*]

Department of Electrical and Computer Engineering, National University of Singapore, 117576 Singapore

We propose two approaches to cloak the spin waves (magnons) by investigating magnetization dynamics. One approach is based on a spatially inhomogeneous anisotropic magnetic moment tensor. The other mechanism is using a spatially inhomogeneous anisotropic gyromagnetic factor tensor and an inhomogeneous external magnetic field. For both approaches, the damping tensor is also inhomogeneous and anisotropic. The magnetic characteristic functions of the magnetic materials have been theoretically derived for both mechanisms. A non-magnetic core, which prevents magnons from entering and consequently distorts the spin wave propagation, can be cloaked by a structured magnetic shell to redirect the spin wave around the core using the above design mechanisms. We discuss the feasibility of the proposed mechanisms in an ensemble of quantum dot molecules and magnetic semiconductors. The proposed approaches shed light on transformation magnonics, and can be utilized for future spin-wave lenses, concentrators, low back-scattering waveguides, and ultimately quantum computing.

[*]eleyang@nus.edu.sg

# I. INTRODUCTION

Invisibility cloaking of different types of waves have been pursued intensively in the last decade.[1-13] Different mechanisms for decreasing the scattering of objects with electromagnetic (EM) waves have been proposed and investigated.[1,3,6,7,13] Amongst those, the space transformation method intuitively provides a promising avenue to achieve the invisibility by designing a particular shell to hide the core. The Maxwell's equations are invariant under such a space transformation, leading to materials in the shell region being inhomogeneous and anisotropic.[6,11] This method or its simplified approximation has been demonstrated theoretically and experimentally for microwave EM waves by using EM metamaterials.[3,6,11-13] It has also been revealed that the two-dimensional Schrodinger equation can be invariant under space transformation resulting in inhomogeneous effective mass and magnetic potential in the shell area.[9] Although it has been suggested that the perfect invariance for general three-dimensional elastodynamic waves is not possible,[5,14] the transformation based cloaking of such waves has been demonstrated for special cases. It has been also proven that the acoustic wave equations remain invariant in 2D and 3D by having a specific inhomogeneity and anisotropy in the mass density and the bulk moduli.[8,15,16] Recently, there have been demonstrations of bilayer cloaks for temperature, dc current, and magnetostatic fields by utilizing bulk homogenous materials.[4,10,17-19] The conception of invisibility is not always reducing the wave scattering. It can be also referred to as hiding a feature of an object or space. As an example, it was proposed that in bilayer graphene, the confined states in the barrier can be cloaked due to chirality mismatch with the continuum states, for the normal and oblique incident currents.[20] Another example is cloaking polarizable discrete systems using an anomalous resonance technique.[7]

Another important type of wave that has been the subject studied for decades, is spin wave with magnon as its quanta.[21] The feature that makes spin waves interesting for applications is their wave-vectors as small as few nanometers and frequencies of tens of GHz. As it is possible to engineer different dispersions in magnetic lattices, the magnonics has thereby attracted a lot of attention as a promising candidate with low energy consumption and high throughput computation, which may possibly go beyond photonic or even electron devices. There have been many theoretical and experimental demonstrations of different types of passive and active magnonic systems and crystals, such as transistors, interferometers, waveguides, and logic gates.[21-24] In addition, due to the high nonlinearity in magnetization dynamics, spin waves have become a base for studying different phenomena such as time reversal and Bose condensation of magnons as Bosonic waves.[25,26]

As the transformation optics enabled new possibilities in photonics, the transformation magnonics can introduce new approaches in magnonics. Despite that a negative refraction index for spin waves[27] has been reported, there has not been a unified account of transformation magnonics for more sophisticated applications such as magnonic invisibility cloaks. As an alternative to the transformation techniques for controlling waves, recent proposals and demonstrations of topologically protected edge modes for photons, phonons, and magnons can be utilized to cloak the defects in the edges.[28-31]

In this work, we investigate the invariance in the governing equations of the magnetization dynamics under the space transformation. The objective is to reduce the scattering of a non-magnetic core (blocking the magnons) by designing the magnetic characteristics in a surrounding shell, such as the magnetic moment, gyromagnetic factor, exchange constant, Gilbert damping, and external magnetic field. The spatial profiles of these parameters in the shell area can be designed such that the magnetization dynamics is rendered maximally invariant under the space transformation.

## II. CLOAKING SHELL DESIGNS BASED ON TRANSFORMATION MAGNONICS

### A. Governing equations of the magnons and possible cloaking approaches

The Landau-Lifshitz-Gilbert (LLG) equation governs the magnetization dynamics. We can write the total magnetization $\vec{M}(\vec{\rho})$ as $\vec{M}(\vec{\rho}) = \vec{M}_0(\vec{\rho}) + \vec{M}_d(\vec{\rho})$, where $\vec{\rho}$ is the coordination vector, $\vec{M}_0(\vec{\rho})$ is the static part of the magnetization, and $\vec{M}_d(\vec{\rho})$ is the dynamic part of the magnetization. For media with isotropic magnetic moments, under the assumption of zero temperature, $|\vec{M}(\vec{\rho})|$ is temporally constant, implying $\vec{M}_0(\vec{\rho}) \cdot \vec{M}_d(\vec{\rho}) = 0$. More generally, we can write $\vec{M}(\vec{\rho}) = \bar{M}_s(\vec{\rho})(\vec{m}_0(\vec{\rho}) + \vec{m}_d(\vec{\rho}))$, where $\bar{M}_s(\vec{\rho})$ is the spatial function of the saturation magnetization tensor, while $\vec{m}_0(\vec{\rho})$ and $\vec{m}_d(\vec{\rho})$ are normalized direction vectors of $\vec{M}_0(\vec{\rho})$ and $\vec{M}_d(\vec{\rho})$, respectively ($\vec{m}_0(\vec{\rho})$ ($\vec{m}_d(\vec{\rho})$) is not necessarily parallel to $\vec{M}_0(\vec{\rho})$ ($\vec{M}_d(\vec{\rho})$)). LLG equation can be written as

$$\bar{M}_s(\vec{\rho}) \frac{\partial \vec{m}(\vec{\rho})}{\partial t} = \bar{\gamma}(\vec{\rho})\left(\bar{M}_s(\vec{\rho})\vec{m}(\vec{\rho}) \times \vec{H}(\vec{\rho})\right) + \bar{\alpha}(\vec{\rho})\bar{\gamma}(\vec{\rho})\left(\bar{M}_s^{-1}(\vec{\rho})\vec{M}_0(\vec{\rho}) \times \left(\bar{M}_s(\vec{\rho})\vec{m}(\vec{\rho}) \times \vec{H}(\vec{\rho})\right)\right), \quad (1)$$

where $\bar{\gamma}(\vec{\rho})$ is the gyromagnetic factor tensor and $\vec{H}(\vec{\rho})$ is the magnetic field that can be written as $\vec{H}(\vec{\rho}) = \vec{H}_{ext}(\vec{\rho}) + \vec{H}_{ex}(\vec{\rho}) + \vec{H}_m(\vec{\rho})$, while $\vec{H}_{ext}(\vec{\rho})$ is the external dc field, and $\vec{H}_{ex}(\vec{\rho})$ and $\vec{H}_m(\vec{\rho})$ are the exchange and dipolar fields induced by $\vec{M}(\vec{\rho})$, respectively. In the equilibrium, $\vec{M}_0(\vec{\rho}) \times \vec{H}_0(\vec{\rho}) = 0$, where $\vec{H}_0(\vec{\rho})$ is the static part of $\vec{H}(\vec{\rho})$. $\vec{H}_{0,ex}$ ($\vec{h}_{d,ex}$) and $\vec{H}_{0,m}$ ($\vec{h}_{d,m}$) are the exchange and dipolar fields arisen from $\vec{M}_0(\vec{\rho})$ ($\vec{M}_d(\vec{\rho})$), respectively. $\vec{m}(\vec{\rho}) = \vec{m}_0(\vec{\rho}) + \vec{m}_d(\vec{\rho})$, and $\bar{\alpha}(\vec{\rho})$ is the Gilbert damping constant tensor.

As a proof-of-concept demonstration, we focus on a cylindrical magnonic cloak. In order to assume dynamic invariance in the out-of-plane direction (i.e., $z$ direction), the thickness of the thin film should be small such that the excitation frequency does not induce modes with out-of-plane wavenumbers. The cylindrical coordinates $r\varphi z$ is transformed into $r'\varphi'z'$, where $r' = g(r)$, $\varphi' = \varphi$, and $z' = z$, indicating that the mapping only occurs to the radial axis. The transformation function should satisfy the boundary conditions of $g(0) = c$ and $g(b) = b$, where $c$ is the core radius, and $b$ is the outer shell radius.

Under such a transformation, we can obtain $\vec{M}_d(\vec{\rho}) = T\vec{M}'_d(\vec{\rho}')$, where $T$ is a diagonal matrix with components of $T_{rr} = \frac{\partial g(r)}{\partial r} = g'(r)$, $T_{\varphi\varphi} = \frac{r'}{r} = \frac{g(r)}{r}$, and $T_{zz} = 1$.[6,11] There are two types of spin waves in terms of propagation, $\vec{q} \parallel \vec{M}_0$ and $\vec{q} \perp \vec{M}_0$, where $\vec{q}$ is the wave-vector. Figure 1(a) shows the schematic vector plot of $\vec{M}_d(\vec{\rho})$, where a cylindrical core is perfectly cloaked from propagating magnons of $\vec{q} \perp \vec{M}_0$ type.

Equation (1) indicates that the magnetization dynamic has dual characteristics, which provide two degrees of freedom for controlling its inertia. One is the magnetic moment which is the manifestation of carrier and orbital spin population in a preferential direction in the Hilbert space ($\vec{M}$) and is determined by the saturation magnetization tensor $\bar{M}_s(\vec{\rho})$. The other one is the factor which determines the modification type and strength of gyration under an application of magnetic fields ($\bar{\gamma}$). Conceptually, with any change in $\bar{M}_s$ or $\bar{\gamma}$, we can modify the point-wise dynamics, as it can be inferred from Eq. (1), and as schematically shown in Fig. 1(b) and (c). Figure 1(b) shows an isotropic magnetization vector ($\vec{M}$) precessing in an anisotropic $\bar{\gamma}$. Figure 1(c) shows an anisotropic magnetization vector ($\vec{M}$) precessing in an isotropic $\bar{\gamma}$. However, we should note that there are two strong non-local correlations in magnetic systems, exchange and dipolar interactions ($\vec{H}_{ex}(\vec{\rho})$ and $\vec{H}_m(\vec{\rho})$), which determine the dispersion relation of the spin waves for given $\bar{M}_s$ and $\bar{\gamma}$. First, we demonstrate a cloaking mechanism based on $\bar{\gamma}$. Subsequently, we utilize $\bar{M}_s$ to cloak the cylindrical core. Finally, we discuss the feasibility of the proposed methods.

Based on the linear perturbation of Eq. (1), we can write the dynamic part of the magnetization as

$$\bar{M}_s(\vec{\rho})\frac{\partial \vec{m}_d(\vec{\rho})}{\partial t} = (\vec{\Pi} + \vec{\Omega}),$$

$$\vec{\Pi}(\vec{\rho}) = \bar{\gamma}(\vec{\rho})\left(\vec{M}_0(\vec{\rho}) \times \vec{h}_d(\vec{\rho}) + \bar{M}_s(\vec{\rho})\vec{m}_d(\vec{\rho}) \times \vec{H}_0(\vec{\rho})\right),$$

$$\vec{\Omega}(\vec{\rho}) = \bar{\alpha}(\vec{\rho})\bar{\gamma}(\vec{\rho})\left(\bar{M}_s^{-1}(\vec{\rho})\vec{M}_0(\vec{\rho}) \times \bar{\gamma}^{-1}(\vec{\rho})\vec{\Pi}(\vec{\rho})\right) \qquad (2)$$

where $\vec{h}_d(\vec{\rho}) = \vec{h}_{d,ex}(\vec{\rho}) + \vec{h}_{d,m}(\vec{\rho})$. Here, we consider the case of the propagating magnons of type $\vec{q} \perp \vec{M}_0$ (the similar discussions apply for $\vec{q} \parallel \vec{M}_0$), while outside the cloaking area ($r > c$) we assume homogeneous and isotropic magnetic moments $\bar{M}_s(\vec{\rho}) = M_{s,0}I$ ($I$ is the identity matrix), and a gyromagnetic factor $\bar{\gamma}(\vec{\rho}) = \gamma_0 I$, where $\vec{M}_0 = M_{s,0}\hat{y}$. We assume $W, L \to \infty$ and $d/W \ll 1$, where $W$ is the width, $L$ is the length, and $d$ is the thickness of the structure (refer to Fig. 1(a)). Under this assumption, the dynamic part of the magnetization has the form $\vec{M}_d(\vec{\rho}) = \vec{M}_{d,0}(\vec{\rho})e^{-ik_x x}$ ($k_x$ is the wavenumber in the $x$ direction), and the dynamic demagnetization tensor $\bar{N}_d$ has only two non-zero components in the $xyz$ coordination system, $N_{d,xx} = -\left[1 - \left(1 - e^{-k_x d}\right)\right]/k_x d$ and $N_{d,zz} = -1 - N_{d,xx}$.[32,33] The dynamic dipolar field is related to $\vec{M}_d(\vec{\rho})$ as $\vec{h}_{d,m}(\vec{\rho}) = \bar{N}_d \vec{M}_d(\vec{\rho})$. In addition, the static demagnetization tensor $\bar{N}_0$ has only one non-zero component $N_{d,zz} = -1$, while $\vec{H}_{0,m}(\vec{\rho}) = \bar{N}_0 \vec{M}_0(\vec{\rho})$. The isotropic exchange field in the continuum limit can be written as $\vec{H}_{ex}(\vec{\rho}) = \Lambda \nabla^2 (\vec{m}_0(\vec{\rho}) + \vec{m}_d(\vec{\rho}))$, where $\Lambda = A/2\pi M_{s,0}^2$ and $A$ is the exchange constant. $\vec{h}_{d,ex}(\vec{\rho}) = -k_x^2 \vec{M}_d(\vec{\rho})$, while the homogeneity of $\vec{M}_0$ implies $\vec{H}_{0,ex}(\vec{\rho}) = 0$.

## B. Cloaking shell designs based on anisotropic inhomogeneous $\bar{\gamma}$

In order to achieve cloaking, Eq. (2) should remain invariant if we rewrite $\vec{M}_d(\vec{\rho})$ as $T\vec{M}'_d(\vec{\rho}')$. The conditions for Eq. (2) to remain invariant in the transformed space are $T^{-1}\vec{\Pi}(\vec{\rho}') = \vec{\Pi}'$ and $T^{-1}\vec{\Omega}(\vec{\rho}') = \vec{\Omega}'$, where

$$\vec{\Pi}' = \bar{\gamma}'(\vec{\rho})\left(\vec{M}'_0(\vec{\rho}') \times \vec{h}'_d(\vec{\rho}') + M_{s,0}\vec{m}'_d(\vec{\rho}') \times \vec{H}'_0(\vec{\rho}')\right) \text{ and}$$

$$\vec{\Omega}' = \bar{\alpha}'(\vec{\rho}')\bar{\gamma}'(\vec{\rho}')\left(\frac{1}{M_{s,0}}\vec{M}'_0(\vec{\rho}') \times \left(\gamma_0^{-1}\vec{\Pi}'\right)\right) \text{ (values with prime refer to the transformed space).}$$

In the $\bar{\gamma}$-based design, we assume that the saturation magnetization is isotropic and $\vec{M}'_s(\vec{\rho}')$ can be replaced by the scalar $M'_{s,0}(\vec{\rho}')$. The invariant conditions to satisfy $T^{-1}\vec{\Pi}(\vec{\rho}') = \vec{\Pi}'$ and $T^{-1}\vec{\Omega}(\vec{\rho}') = \vec{\Omega}'$ are derived to be

$$H'_{0,\varphi}(\vec{\rho}') = T_{rr}H_{0,\varphi}(\vec{\rho}'), \quad M'_{0,\varphi}(\vec{\rho}') = T_{rr}M_{0,\varphi}(\vec{\rho}'),$$

$$H'_{0,r}(\vec{\rho}') = T_{\varphi\varphi}H_{0,r}(\vec{\rho}'), \quad M'_{0,r}(\vec{\rho}') = T_{\varphi\varphi}M_{0,r}(\vec{\rho}'),$$

$$\gamma'_{rr}(\vec{\rho}') = \gamma_0 \frac{1}{T_{rr}^2}, \quad \gamma'_{\varphi\varphi}(\vec{\rho}') = \gamma_0 \frac{1}{T_{\varphi\varphi}^2}, \quad \gamma'_{zz} = \gamma_0,$$

$$\alpha'_{rr(\varphi\varphi)}(\vec{\rho}') = \alpha_0, \quad \alpha'_{zz}(\vec{\rho}') = \alpha_0 \frac{(M_{0,r}(\vec{\rho}'))^2 + (M_{0,\varphi}(\vec{\rho}'))^2}{T_{\varphi\varphi}^2(M_{0,r}(\vec{\rho}'))^2 + T_{rr}^2(M_{0,\varphi}(\vec{\rho}'))^2}. \quad (3)$$

$\gamma'_{ii}$ ($\alpha'_{ii}$), $i = r, \varphi$, and z, are the diagonal components of $\bar{\gamma}'$ ($\bar{\alpha}'$). $\alpha_0$ is the value of the homogeneous Gilbert damping constant outside the cloaking area ($r > c$). The steps to achieve the invariant conditions presented in Eq. (3) are explained in Appendix. It should be mentioned that the conditions in Eq. (3) satisfy the invariant conditions in all three directions of the cylindrical coordination system, if $\vec{h}_d(\vec{\rho}) = T\vec{h}'_d(\vec{\rho}')$ (note that $\vec{M}_d(\vec{\rho}) = T\vec{M}'_d(\vec{\rho}')$). This condition holds, because $\vec{h}_d(\vec{\rho})$ can be written as a linear function of $\vec{M}_d(\vec{\rho})$ ($\vec{h}_d(\vec{\rho}) = (\bar{N}_d - k_x^2)\vec{M}_d(\vec{\rho})$, where $\bar{N}_d$ does not have a spatial functionality) under the assumptions given above ($\bar{M}_s(\vec{\rho}) = M_{s,0} I, W, L \to \infty, d/W \ll 1$, and $\vec{M}_d(\vec{\rho}) = \vec{M}_{d,0}(\vec{\rho}) e^{-ik_x x}$).

The functionality of $M'_{0,\varphi(r)}(\vec{\rho}')$ in Eq. (3) will modify the static field $H_{0,\varphi(r)}(\vec{\rho}')$ through the exchange and dipolar fields, which will contradict the other assumptions which led us to the established invariance conditions. To cancel the redundant static exchange and dipolar fields, $H_{0,\varphi(r)}(\vec{\rho}')$ should be modified as

$$H_{0,\varphi(r)}(\vec{\rho}') = H'_{0,\varphi(r)}(\vec{\rho}') + \left[ -\left( \int_V \bar{G}(\vec{\rho}', \vec{\tau}) \vec{M}'_0(\vec{\tau}) d\tau \right) - \Lambda \nabla^2 \left( \vec{m}'_0(\vec{\rho}') \right) \right] \cdot \hat{\varphi}(\hat{r}), \qquad (4)$$

where $\bar{G}(\vec{\rho}', \vec{\tau})$ is the dipolar Green function tensor in the cylindrical coordination system. Therefore, the perfect cloaking of magnons which are governed by Eq. (2) can be achieved for the aforementioned assumptions, if Eq. (3-4) are satisfied. We name this method the $\bar{\gamma}$-mechanism, contrasting the $\bar{M}_s$-mechanism that will be described later. It can be inferred from Eq. (3), that in the $\bar{\gamma}$-mechanism, $\bar{M}'_s(\vec{\rho}') = M'_s(\vec{\rho}') I$, where $M'_s(\vec{\rho}') = |\vec{M}'_0(\vec{\rho}')|$ is not homogeneous but is isotropic.

## C. Cloaking shell designs based on anisotropic inhomogeneous $\vec{M}_s$

For the $\bar{\gamma}$-mechanism, we investigate the conditions that should be held for Eq. (2) to remain invariant under the space transformation required for cloaking ($T^{-1}\vec{\Pi}(\vec{\rho}') = \vec{\Pi}'$ and $T^{-1}\vec{\Omega}(\vec{\rho}') = \vec{\Omega}'$). However, the magnetization is a vector field whose magnitude can be anisotropic, enabling another approach for rendering Eq. (2) invariant. In the transformed space, the dynamic part of the magnetization should be $\vec{M}'_d(\vec{\rho}') = T^{-1}\vec{M}_d(\vec{\rho})$, which can be achieved if

$$\bar{M}'_s(\vec{\rho}') = M_{s,0}T^{-1}. \tag{5}$$

The conditions to achieve invariance in Eq. (2) with the assumption of Eq. (5), become (refer to Appendix for details)

$$H'_{0,\varphi}(\vec{\rho}') = H_{0,\varphi}(\vec{\rho}'),\ M'_{0,\varphi}(\vec{\rho}') = M_{0,\varphi}(\vec{\rho}'),\ m'_{0,\varphi}(\vec{\rho}') = T_{\varphi\varphi}m_{0,\varphi}(\vec{\rho}'),$$

$$H'_{0,r}(\vec{\rho}') = H_{0,r}(\vec{\rho}'),\ M'_{0,r}(\vec{\rho}') = M_{0,r}(\vec{\rho}'),\ m'_{0,r}(\vec{\rho}') = T_{rr}m_{0,r}(\vec{\rho}'),$$

$$\alpha'_{rr}(\vec{\rho}') = \frac{\alpha_0}{T_{\varphi\varphi}},\ \alpha'_{\varphi\varphi}(\vec{\rho}') = \frac{\alpha_0}{T_{rr}},\ \alpha'_{zz}(\vec{\rho}') = \alpha_0 \frac{\left(M_{0,r}(\vec{\rho}')\right)^2 + \left(M_{0,\varphi}(\vec{\rho}')\right)^2}{T_{rr}\left(M_{0,r}(\vec{\rho}')\right)^2 + T_{\varphi\varphi}\left(M_{0,\varphi}(\vec{\rho}')\right)^2}. \tag{6}$$

It should be noted that the gyromagnetic factor is assumed to be isotropic for the $\bar{M}_s$-mechanism, $\bar{\gamma}'(\vec{\rho}') = \gamma_0 I$. In contrast to the $\bar{\gamma}$-mechanism, $M'_{0,r(\varphi)}(\vec{\rho}') = M_{0,r(\varphi)}(\vec{\rho}')$ holds for the $\bar{M}_s$-mechanism (comparing Eq. (3) and Eq. (6)), therefore there is no modification in the static exchange or dipolar field ($H_{0,\varphi(r)}(\vec{\rho}') = H'_{0,\varphi(r)}(\vec{\rho}')$).

# III. NUMERICAL DEMONSTRATION OF THE CLOAKING MECHANISMS

In order to demonstrate the functionality of the proposed magnonic cloak, we have developed an in-house code to solve the LLG equation (Eq. (1)) for the anisotropic $\bar{\gamma}$, $\bar{M}_s$, and $\bar{\alpha}$.[34] We assume the structure dimensions in Fig. 1(a) to be $W = 420$ nm, $L = 820$ nm, $c = 50$ nm, $b = 100$ nm, $d = 3.8$ nm, $x_c = 550$ nm, and $y_c = 210$ nm ($x_c$ and $y_c$ are the center positions of the cylindrical core along the $x$ and $y$ direction, respectively, with respect to the origin). The transformation function $g(r) = c + r(b-c)/b$ is employed. The meshing cells are cubic and have the dimensions of 2 nm × 2 nm × 3.8 nm. In the non-transformed area, we assume $\vec{m}_0 = \hat{y}$, $\vec{H}_{ext} = 10000\hat{y}$ Oe, $M_{s,0} = 8 \times 10^5$ A/m, $A = 0.5 \times 10^{-11}$ J/m, and $\gamma_0 = 2.2 \times 10^5$ Hz/(A·m). The microwave excitation is applied as $\vec{h}_{mw} = 1 \times sin(\omega_{mw}t)\hat{x}$ Oe, at $x = 800$ nm. The microwave excitation frequency was set as $\omega_{mw} = 2\pi \times 50 \times 10^9$ rad/s. We apply matched layers (ML) of 4 nm width in all four in-plane boundaries. In the ML area $\alpha_0 = 1$, while $\alpha_0 = 0.01$ for the rest of the structure.

Figure 2 shows the snapshot of the magnetization in the $z$ direction ($M_z$) at $t = 2.2$ ns (well before the wave reaches the ML layer at $x = 0$, in which case the reflection drives its adjacent magnetization dynamics unstable). The dashed lines at $x = 800$ nm represent the microwave excitation lines, inducing spin waves propagating in the $-x$ direction. The horizontal lines in Fig. 2 separate regions of each graph that have a specific color-code shown in their right side. Figure 2(a) demonstrates the spin wave configuration when there is no cylindrical core ($b = 0$ and $c = 0$). It can be observed that due to the finite width ($W = 420$ nm), in addition to the propagating magnons in the $-x$ direction, standing spin waves are formed across the $y$ direction ($k_y \neq 0$). Figure 2(b) shows a case with a cylindrical core while no cloaking mechanism was applied ($b = 50$ nm and $c = 50$ nm). The shadowing of the core in the spin wave configuration can be clearly observed in Fig. 2(b). Figure 2(c) and Figure 2(d) show the cases where $\bar{\gamma}$-mechanism and $\bar{M}_s$-mechanism were applied, respectively, with $b = 100$ nm and $c = 50$ nm. Both Fig. 2(c) and Fig. 2(d) demonstrate the reduction of the shadow of the core in comparison with Fig. 2(b).

The values of $M_{z,sr}(x) = \dfrac{1}{20\,nm} \displaystyle\int_{y_c-10nm}^{y_c+10nm} M_z(x,y)\,dy$ are shown in Fig. 3(a) and 3(b) for $0 \leq x \leq 420$ nm and $0 \leq x \leq 820$ nm, respectively. Especially, Fig. 3(a) shows a quantitative comparison of $M_z$ in the shadowing region (the dashed boxes in Fig. 2) for all the four cases in Fig. 2. It can be seen that the spin-waves are suppressed after the propagation through the core where no cloaking mechanism is applied, while both the cloaking mechanisms have the values of $M_{z,sr}(x)$ close to that of no core. Despite this signature of cloaking (reduction in the core shadow) shown in Fig. (2) and Fig. 3(a), Fig. 3(c), showing that the total average of $M_z$ in the $y$ direction ($M_{z,T}(x) = \dfrac{1}{420\,nm} \displaystyle\int_{0nm}^{420nm} M_z(x,y)\,dy$) for $0 \leq x \leq 420$ nm, has almost the same value for all four cases. The reason is that the energy exchange between the standing spin wave in the $y$ direction and the propagating spin wave in the $x$ direction, provides a nonlinear route for the magnon population to pass through such cores. This can be also justified by the results in Fig. 3(b) and (d) that show $M_{z,sr}(x)$ and $M_{z,T}(x)$ for the whole range of $x$, respectively, with similar amplitudes for all four cases. If such standing spin waves in the $y$ direction are omitted from the system by expanding the width of the magnetic structure ($W \to \infty$), we can expect to observe higher reflection and shadow of the core with no cloaking mechanism, and more clear cloaking for both the $\bar{\gamma}$-mechanism and $\bar{M}_s$-mechanism.

The main reason behind the imperfection of the cloaking for both of the mechanisms (refer to Fig 2(c-d)) is that due to the existence of the waveform in the y direction, the assumption of $\vec{h}_d(\vec{\rho}) = T\vec{h}'_d(\vec{\rho}')$, which was the basis for derivation of the material properties in the shell, is no longer perfectly satisfied. $\vec{h}_d(\vec{\rho}) = T\vec{h}'_d(\vec{\rho}')$ holds if only one of the $\vec{q} \perp \vec{M}_0$ or $\vec{q} \parallel \vec{M}_0$ modes exists. Other reasons for the cloaking imperfections observed in Fig. 2(c-d) are using cubic (aligned with the Cartesian axes) and limited number of cells in the shell for the simulations. The difference between the results of the $\bar{\gamma}$-mechanism and $\bar{M}_s$-mechanism, corresponding to Fig. 2(c) and (d) respectively, originates from the degree of vulnerability of the methods with respect to the discrepancy of $\vec{h}_d(\vec{\rho}) = T\vec{h}'_d(\vec{\rho}')$ from perfection due to mixing of the $\vec{q} \perp \vec{M}_0$ and $\vec{q} \parallel \vec{M}_0$ modes. For the $\bar{M}_s$-mechanism, mixing affects the dynamic dipolar field ($\vec{h}'_{d,m}(\vec{\rho}')$) leading to distortion and inaccuracy of Eq. (6), while for the $\bar{\gamma}$-mechanism, in addition to the distortion induced by $\vec{h}'_{d,m}(\vec{\rho}')$, Eq. (3) no longer holds exactly. Therefore, more distortion for the $\bar{\gamma}$-mechanism in comparison with the $\bar{M}_s$-mechanism in the presence of mode-mixing is expected, as inferred by comparing Fig. 2(c) and Fig. 2(d).

## IV. SPIN-METAMATERIALS FOR TRANSFORMATION MAGNONICS

### A. Physical feasibility of the $\bar{\gamma}$-mechanism

Figures 4(a-c) show the material properties ($\gamma_{rr}/\gamma_{\varphi\varphi}, M_s(\vec{\rho})$, and $\alpha_{zz}/\alpha_0$) and the direction of the external field $\vec{H}_{ext}(\vec{\rho})$ for the $\bar{\gamma}$-mechanism based on Eq. (3). Figure 4(e-f)

show $M_{s,rr}/M_{s,\varphi\varphi}$ and $\alpha_{zz}/\alpha_0$ for the $\bar{M}_s$-mechanism based on Eq. (5) and (6). Magnon cloaking by the $\bar{\gamma}$-mechanism or the $\bar{M}_s$-mechanism cannot be achieved in metallic ferromagnets as the large anisotropy in $\bar{\gamma}$ or $\bar{M}_s$ is not possible. In addition, the anisotropy and tuning range of $\bar{\gamma}$ is limited in the magnetic molecules or magnetic semiconductors for realizing the $\bar{\gamma}$-mechanism (refer to Fig. 4(a)).[35,36] However, high anisotropy and large tuning range of $\bar{\gamma}$ is possible in quantum dot molecules (QDM) (refer to the box in Fig. 4(d) for a stacked quantum dot molecule schematic).[37-40] Spin states in quantum dots (QD) or QDMs are the main candidates for quantum computing.[41] It has been demonstrated theoretically and experimentally that the Lande g-factor or $\bar{\gamma}$ in our notation, can be tuned in a large range (including zero crossing) in quantum wells (QW), QDs, and QDMs with an electric field. However for QDMs, the effect of electric field is much richer on tuning both the amplitude and the anisotropy of the hole-spin Lande g-factor. There are rich crossings and anti-crossings for the ground, the excitonic, and the charged excitonic states.[37-39,42-48] The crossings and anti-crossings in QDMs happen due to bonding and anti-bonding of electron and hole wave functions between the two QDs in a typical stacked QDM (refer to Fig. 4(d)) which occur by changing the electric field.[37,39,42] In the right part of the Fig. 4(d), a schematic of a QDM, its respective Cartesian coordination, the applied electric field in the $z$ direction ($E_{z,q}$, where $q$ is the number of the QDM in the QDM ensemble) and the magnetic field $\vec{B}_q$ are depicted.

It has been demonstrated theoretically that for a QDM in Fig. 4(d), the axes of the Lande g-factor ($\bar{\gamma}$) ellipsoid are along the $z$ direction, $\hat{x}_q + \hat{y}_q$, and $\hat{x}_q - \hat{y}_q$ ($x_q$ and $y_q$ are the local Cartesian direction for the QDM number $q$). Based on this information about QDMs, we propose utilizing the $\bar{\gamma}$-mechanism for an ensemble of QDMs as depicted schematically in the left part of Fig. 4(d). If we assume $\left(\hat{x}_q+\hat{y}_q\right)\|\hat{r}$ and $\left(\hat{x}_q-\hat{y}_q\right)\|\hat{\varphi}$, the value of $\gamma_{zz}$ can be tuned for more than 100%, while $\gamma_{rr}$ and $\gamma_{\varphi\varphi}$ can be tuned for up to 800% with varying $E_{z,q}$.[39] However, $\gamma_{rr}$ and $\gamma_{zz}$ are spatially constant in the shell, while varying $E_{z,q}$ for tuning $\gamma_{\varphi\varphi}$ will change $\gamma_{rr}$ and $\gamma_{zz}$ as well. To overcome this issue, we can utilize the local angle ($\theta_q$) of $\hat{x}_q+\hat{y}_q$ with respect to $\hat{r}$ as another variable. In addition, the application of a spatially functionalized strain (adjacent piezoelectric layers) or doping can be used as other tuning factors for achieving the desired $\bar{\gamma}$ at the position of QDM number $q$ (refer to Fig. 4(a)). To achieve the required $M_s(\vec{\rho})$ configuration for the $\bar{\gamma}$-mechanism (refer to Fig. 4(b)), the density of the QDMs in the cloaking shell should be spatially functionalized. However, the inhomogeneous distribution of QDMs in the shell causes inhomogeneity of distance between the QDMs which directly affects the respective exchange mechanisms. To compensate this effect, a spatially functionalized electric field can be applied in the semiconductor regions between the QDMs to tune the exchange strength.[49,50] The spin lifetime in QDs can be up to the order of μs,[51] and due to the atomic like behavior of QDs, the phenomenological Gilbert damping and the required $\alpha_{zz}/\alpha_0$ configuration (refer to Fig. 4(c)) can be ignored.

## B. Physical feasibility of the $\bar{M}_s$-mechanism

There are theoretical and experimental demonstrations for anisotropies in both the moment and exchange interaction in magnetic semiconductors.[52-62] Those anisotropies stem from the spin orbit interactions (SOI) in the materials that lack the inversion symmetry. In bulk semiconductors with wurtzite or zinc-blende crystalline structures, the antisymmetric part of the anisotropic exchange of the localized electrons is dominated by Dzyaloshinskii-Moriya interaction (DMI)[63,64] which is the first-order perturbation in SOI of Rashba[65] and Dyakonov-Kachorovskii[66] types. There have been experimental demonstrations of anisotropic exchange by showing anisotropic dephasing in bulk GaN and impurity-bound electrons in n-doped ZnO.[54,55]

There have been theoretical and experimental demonstrations of magnetic orderings in semiconductors.[52,56,60,67-83] Such magnetic orderings have been achieved due to the presence of carrier-doping, cation vacancy, cation substitution, anion vacancy, anion substitution, interstitial impurities, structural strain, and the combination of them. Although a wide range of doping and defect gives rise to local spins or orbital moments, not all of them forms a long range magnetic order. The exchange interaction between local moments is governed by several mechanisms, such as double exchange and superexchange. In addition to large magnetic orderings, the anisotropy in moments has been demonstrated in several oxides like substituted ZnO, V doped $SnO_2$, $HfO_2$, and $TiO_2$, as well as $Li_2(Li_{1-x}Fe_x)N$.[52,53,56,60,69,83,84] The anisotropic moment arises due to lifting the degeneracy in orbital interactions, and mixing of molecular orbitals surrounding the point defects induced by oxygen vacancies, cation vacancies and interstitial or substitution impurities. Molecular orbitals surrounding the point defects mix with the nearest neighbors and next nearest neighbors (source of magnetic ordering), enabling the possibility of different anisotropy patterns based on the respective position of the impurities (depends on the SOI strength of the defect).

In order to be more specific, we propose a system of interstitial impurities and oxygen vacancies in a metal-oxide. Figure 4(g) shows a 2×2×1 supercell of rutile crystalline structure consisting of metal sites and oxygen sites, hosting two interstitial impurities and an oxygen vacancy, for example. The surrounding oxygen octahedral of the interstitial imposes a crystal field on the impurity and possibly splits the degenerate bands based on the symmetry rules. If the bonding molecular orbitals induced by impurities are filled with carriers, an orbital moment can be generated. In the presence of spin-orbit coupling, the impurity induced spin-moment aligns with the orbital moment and if an exchange mechanism exists, both the orbital moment and spin moment can give rise to a macroscopic ferromagnetic order. The presence of oxygen vacancy ($V_o$) provides electrons, and if its defect state overlaps with the impurity induced bands, there could be both orbital moments and long-range exchange interaction.[52,56,60,68,73,75,76,79,80,82] Therefore, the choice of materials for the host metal and the interstitial metal is important to have the required interactions. The amplitude and the direction of the electric field $\vec{E}_i$ on the supercell $i$ can determine the respective configuration of the impurities and defects.

In the rutile structure, there are four independent octahedral sites for interstitial impurities as indicated in Fig. 4(h). Figure 4(h) shows a simplified demonstration of the molecular orbitals as charge rings $\sigma_{1(2,3,4)}$ that give rise to orbital moment vectors $\vec{v}_{1(2,3,4)}$. The coexistence of orbital moment and high SOC results in anisotropic moments. In order to control the axis of anisotropy, there is a need for at least two interacting interstitial impurities. Figure 4(i) shows a possible route to control the anisotropy axis in the entire three dimensions. If the distance of the $V_o$ to the interstitial impurity in site $A$ is less than its distance to site $B$, the carrier density in the charge ring of $A$ will be higher than that of $B$, therefore the orbital magnetization in $A$ will be higher than that in $B$ ($|\vec{v}_A| > |\vec{v}_B|$). Hybridization of $A$ and $B$ charge rings results in a net orbital moment $\vec{v}_T$. It can be inferred from Fig. 4(h) and 4(i) that the direction of $\vec{v}_T$ can be tuned by placing the impurities in different octahedral sites and placing the oxygen vacancy in different oxygen sites. Spatial functionalization of the temperature,[56] impurity concentration, charge doping, and oxygen vacancy can be utilized for tuning the amplitude of the moment as is needed in addition to the anisotropy direction to achieve the desired $\bar{M}_s$.

The proposed method of transformation magnonics for spin wave cloaking might be very challenging to be realized experimentally, as it requires spatially varying anisotropic $\bar{\gamma}$ or $\bar{M}_s$ with precision in nanometer scale. However, the proposed mechanisms may find plausible applications in simpler transformation designs such as magnon lenses, concentrators, bending waveguides, and ultimately spin based quantum computing.

## V. CONCLUSION

We have proposed two transformation-magnonics based approaches for cloaking of a cylindrical non-magnetic core. The $\bar{\gamma}$-mechanism imposes an inhomogeneous anisotropic gyromagnetic tensor in the cloaking shell, while the $\bar{M}_s$-mechanism is based on inhomogeneous and anisotropic magnetic moments. We show that the wave-front of the incident spin-wave remains invariant after propagating through the shell for both the mechanisms, indicating that the non-magnetic cylindrical core has been invisible towards the incident magnons. We discuss the feasibility of the $\bar{\gamma}$-mechanism in the ensemble of quantum dot molecules. We also propose functionalized defects in magnetic oxides for the feasibility of the $\bar{M}_s$-mechanism. The reported design mechanism of transformation magnonics for manipulating magnons in magnetic semiconductors or quantum dot ensembles paves an alternative way for realizing advanced functionalities such as magnonic cloaking, lensing, and concentrations, etc.

## ACKNOWLEDGEMENTS

This work was supported by the Singapore Ministry of Education Academic Research Fund Tier 1 (R-263-000-A46-112).

## APPENDIX

For the $\bar{\gamma}$-mechanism, $T^{-1}\vec{\Pi}(\vec{\rho}') = \vec{\Pi}'$ results in three coupled equations, which with a straightforward algebraic investigation determine the invariant conditions for different spatial coordinates of $\vec{H}'_0(\vec{\rho}')$ and $\vec{M}'_0(\vec{\rho}')$ based on the matrix components of $T$. Such invariance is achieved using the $\bar{\gamma}'(\vec{\rho}')$ tensor as a degree of freedom, as shown in Eq. (3). To be more specific, the $T^{-1}\vec{\Pi}(\vec{\rho}') = \vec{\Pi}'$ in the $z$ direction leads to (note that $\vec{h}_d(\vec{\rho}) = T\vec{h}'_d(\vec{\rho}')$ and $\vec{M}_d(\vec{\rho}) = T\vec{M}'_d(\vec{\rho}')$)

$$\frac{\gamma_0}{T_{zz}}\left[\left(M_{0,rr}T_{\varphi\varphi}h'_{d,\varphi\varphi} - M_{0,\varphi\varphi}T_{rr}h'_{d,rr}\right) + \left(T_{rr}M'_{d,rr}H_{0,\varphi\varphi} - T_{\varphi\varphi}M'_{d,\varphi\varphi}H_{0,rr}\right)\right]\hat{z} = $$
$$\gamma'_{zz}\left[\left(M'_{0,rr}h'_{d,\varphi\varphi} - M'_{0,\varphi\varphi}h'_{d,rr}\right) + \left(M'_{d,rr}H'_{0,\varphi\varphi} - M'_{d,\varphi\varphi}H'_{0,rr}\right)\right]\hat{z} \quad (A1)$$

With $T_{zz} = 1$, Eq. (A1) can be satisfied if $\gamma'_{zz} = \gamma_0$, $H'_{0,\varphi}(\vec{\rho}') = T_{rr}H_{0,\varphi}(\vec{\rho}')$, $M'_{0,\varphi}(\vec{\rho}') = T_{rr}M_{0,\varphi}(\vec{\rho}')$, $H'_{0,r}(\vec{\rho}') = T_{\varphi\varphi}H_{0,r}(\vec{\rho}')$, and $M'_{0,r}(\vec{\rho}') = T_{\varphi\varphi}M_{0,r}(\vec{\rho}')$. In the presence of the latter conditions and noting that $M_{0,z}(\vec{\rho}') = 0$ and $H_{0,z}(\vec{\rho}') = 0$, the invariance of $T^{-1}\vec{\Pi}(\vec{\rho}') = \vec{\Pi}'$ in the $\varphi$ and $r$ directions can be achieved if

$$\frac{\gamma_0}{T_{rr}}\left[\left(-\frac{M'_{0,\varphi\varphi}}{T_{rr}}T_{zz}h'_{d,zz}\right) + \left(T_{zz}M'_{d,zz}\frac{H'_{0,\varphi\varphi}}{T_{rr}}\right)\right]\hat{r} = \gamma'_{rr}\left[\left(-M'_{0,\varphi\varphi}h'_{d,zz}\right) + \left(M'_{d,zz}H'_{0,\varphi\varphi}\right)\right]\hat{r},$$

$$\frac{\gamma_0}{T_{\varphi\varphi}}\left[\left(\frac{M'_{0,rr}}{T_{\varphi\varphi}}T_{zz}h'_{d,zz}\right) + \left(-T_{zz}M'_{d,zz}\frac{H'_{0,rr}}{T_{\varphi\varphi}}\right)\right]\hat{\varphi} = \gamma'_{\varphi\varphi}\left[\left(M'_{0,rr}h'_{d,zz}\right) + \left(M'_{d,zz}H'_{0,rr}\right)\right]\hat{\varphi}. \quad (A2)$$

Equation (A2) can be satisfied if $\gamma'_{rr}(\vec{\rho}') = \gamma_0 \frac{1}{T_{rr}^2}$ and $\gamma'_{\varphi\varphi}(\vec{\rho}') = \gamma_0 \frac{1}{T_{\varphi\varphi}^2}$, respectively. By using the invariance conditions that satisfy $T^{-1}\vec{\Pi}(\vec{\rho}') = \vec{\Pi}'$, the constraints on the components of the Gilbert damping tensor $\bar{\alpha}'$ are determined by considering $T^{-1}\vec{\Omega}(\vec{\rho}') = \vec{\Omega}'$ with following similar procedures (see Eq. (3) for the $\bar{\alpha}'$ components).

For the $\bar{M}_s$-mechanism, $\vec{\Pi}(\vec{\rho}') = \vec{\Pi}'$ should be satisfied instead of $T^{-1}\vec{\Pi}(\vec{\rho}') = \vec{\Pi}'$, because the assumption in Eq. (5) for the $\bar{M}'_s$ tensor renders the left part of Eq. (2) invariant. To satisfy $\vec{\Pi}(\vec{\rho}') = \vec{\Pi}'$, the condition of Eq. (5) is only required, and the values of the bias field and the static part of the magnetization in the prime and the physical space remain equivalent, $\vec{H}'_0(\vec{\rho}') = \vec{H}_0(\vec{\rho}')$ and $\vec{M}'_0(\vec{\rho}') = \vec{M}_0(\vec{\rho}')$. It should be noted that, since $\bar{M}'_s$ is anisotropic, $\vec{M}'_0(\vec{\rho}')$ is not necessarily parallel to $\vec{m}'_0(\vec{\rho}')$ in the shell area (refer to Eq. (6)). Similar to the $\bar{\gamma}$-mechanism, the invariance of $\vec{\Pi}(\vec{\rho}') = \vec{\Pi}'$ is guaranteed only if $\vec{h}_d(\vec{\rho}) = T\vec{h}'_d(\vec{\rho}')$ and $\vec{M}_d(\vec{\rho}) = T\vec{M}'_d(\vec{\rho}')$, which is achieved under the assumptions stated previously ($\bar{M}_s(\vec{\rho}) = M_{s,0}I$, $W, L \to \infty$, $d/W \ll 1$, and $\vec{M}_d(\vec{\rho}) = \vec{M}_{d,0}(\vec{\rho})e^{-ik_x x}$). Finally, $\vec{\Omega}(\vec{\rho}') = \vec{\Omega}'$ should be also satisfied, which requires that the components of the tensor $\bar{\alpha}'$ to be as indicated in Eq. (6).

**References**

1. A. Alù and N. Engheta, Phys. Rev. E **72,** 016623 (2005).
2. B. Liao, M. Zebarjadi, K. Esfarjani, and G. Chen, Phys. Rev. Lett. **109,** 126806 (2012).
3. D. Schurig, J. J. Mock, B. J. Justice, S. A. Cummer, J. B. Pendry, A. F. Starr, and D. R. Smith, Science **314,** 977 (2006).
4. F. Gömöry, M. Solovyov, J. Šouc, C. Navau, J. P.-Camps, and A. Sanchez, Science **335,** 1466 (2012).
5. G. W. Milton, M. Briane, and J. R. Willis, New J. Phys. **8,** 248 (2006).
6. J. B. Pendry, D. Schurig, and D. R. Smith, Science **312,** 1780 (2006).
7. N.-A. P. Nicorovici, G. W. Milton, R. C. McPhedran, and L. C. Botten, Optic Express **15,** 6314 (2007).
8. S. A. Cummer and D. Schurig, New J. Phys. **9,** 45 (2007).
9. S. Zhang, D. A. Genov, C. Sun, and X. Zhang, Phys. Rev. Lett. **100,** 123002 (2008).
10. T. Han, X. Bai, D. Gao, J. T. L. Thong, B. Li, and C.-W. Qiu, Phys. Rev. Lett. **112,** 054302 (2014).
11. U. Leonhardt and T. G. Philbin, New J. Phys. **8,** 247 (2006).
12. W. Cai, U. K. Chettiar, A. V. Kildishev, and V. M. Shalaev, Nature Photon. **1,** 224 (2007).
13. W. Cai, U. K. Chettiar, A. V. Kildishev, V. M. Shalaev, and G. W. Milton, Appl. Phys. Lett. **91,** 111105 (2007).
14. S. A. Cummer, B.-I. Popa, D. Schurig, D. R. Smith, J. Pendry, M. Rahm, and A. Starr, Phys. Rev. Lett. **100,** 024301 (2008).
15. A. Diatta and S. Guenneau, Appl. Phys. Lett. **105,** 021901 (2014).
16. M. Farhat, S. Guenneau, and S. Enoch, Phys. Rev. B **85,** 020301 (2012).
17. A. Sanchez, C. Navau, J. P.-Camps, and D.-X. Chen, New J. Phys. **13,** 093034 (2011).
18. T. Han, H. Ye, Y. Luo, S. P. Yeo, J. Teng, S. Zhang, and C.-W. Qiu, Adv. Mater. **26,** 3478 (2014).
19. B. Zhang, Y. Luo, X. Liu, and G. Barbastathis, Phys. Rev. Lett. **106,** 033901 (2011).
20. N. Gu, M. Rudner, and L. Levitov, Phys. Rev. Lett. **107,** 156603 (2011).
21. M. Krawczyk and D. Grundler, J. Phys.: Condens. Matter. **26,** 123202 (2014).
22. J. H. Kwon, S. S. Mukherjee, P. Deorani, M. Hayashi, and H. Yang, Appl. Phys. A **111,** 369 (2013).
23. M. Jamali, J. H. Kwon, S.-M. Seo, K.-J. Lee, and H. Yang, Sci. Rep. **3,** 3160 (2013).
24. A. V. Chumak, A. A. Serga, and B. Hillebrands, Nat. Commun. **5,** 4700 (2014).
25. A. V. Chumak, V. S. Tiberkevich, A. D. Karenowska, A. A. Serga, J. F. Gregg, A. N. Slavin, and B. Hillebrands, Nat. Commun. **1,** 141 (2010).
26. S. O. Demokritov, V. E. Demidov, O. Dzyapko, G. A. Melkov, A. A. Serga, B. Hillebrands, and A. N. Slavin, Nature **443,** 430 (2006).
27. S.-K. Kim, S. Choi, K.-S. Lee, D.-S. Han, D.-E. Jung, and Y.-S. Choi, Appl. Phys. Lett. **92,** 212501 (2008).
28. L. Zhang, J. Ren, J.-S. Wang, and B. Li, Phys. Rev. Lett. **105,** 225901 (2010).
29. A. Mook, U. Henk, and I. Mertig, Phys. Rev. B **91,** 174409 (2015).
30. R. Shindou, R. Matsumoto, S. Murakami, and J.-I. Ohe, Phys. Rev. B **87,** 174427 (2013).
31. A. B. Khanikaev, S. H. Mousavi, W.-K. Tse, M. Kargarian, A. H. MacDonald, and G. Shvets, Nature Mater. **12,** 233 (2012).
32. B. A. Kalinikos and A. N. Slavin, J. Phys. C: Solid State Phys. **19,** 7013 (1986).
33. M. P. Kostylev, G. Gubbiotti, J.-G. Hu, G. Carlotti, T. Ono, and R. L. Stamps, Phys. Rev. B **76,** 054422 (2007).
34. In each time step, first the exchange and dipolar fields were calculated. Exchange fields were calculated using only the nearest neighbors. Dipolar fields were calculated using the fast Fourier transform method. After field calculations, the anisotropic LLG differential equation, Eq. (1), multiplied by $\bar{M}_s^{-1}$ in both sides, was solved using the Runge-Kutta numerical method for the magnetization of each cell. The maximum time step of 0.01 ps was used. The time-steps were modified to meet the relative error criteria of $10^{-9}$.

35. A. Trueba, P. Garcia-Fernandez, F. Senn, C. A. Daul, J. A. Aramburu, M. T. Barriuso, and M. Moreno, Phys. Rev. B **81,** 075107 (2010).
36. G. Cucinotta, M. Perfetti, J. Luzon, M. Etienne, P.-E. Car, A. Caneschi, G. Calvez, K. Bernot, and R. Sessoli, Angew. Chem. **51,** 1606 (2012).
37. H. J. Krenner, M. Sabathil, E. C. Clark, A. Kress, D. Schuh, M. Bichler, G. Abstreiter, and J. J. Finley, Phys. Rev. Lett. **94,** 057402 (2005).
38. W. Sheng and J.-P. Leburton, Phys. Rev. Lett. **88,** 167401 (2002).
39. T. Andlauer and P. Vogl, Phys. Rev. B **79,** 045307 (2009).
40. R. Roloff, W. Potz, T. Eissfeller, and P. Vogl, arXiv **1003.0897v1** (2010).
41. D. Loss and D. P. DiVincenzo, Phys. Rev. A **57,** 120 (1998).
42. E. A. Stinaff, M. Scheibner, A. S. Bracker, I. V. Ponomarev, V. L. Korenev, M. E. Ware, M. F. Doty, T. L. Reinecke, and D. Gammon, Science **311,** 636 (2006).
43. Y. Kato, R. C. Myers, D. C. Driscoll, A. C. Gossard, J. Levy, and D. D. Awschalom, Science **299,** 1201 (2003).
44. T. P. Mayer Alegre, F. G. G. Hernandez, A. L. C. Pereira, and G. Medeiros-Ribeiro, Phys. Rev. Lett. **97,** 236402 (2006).
45. J. R. Petta and D. C. Ralph, Phys. Rev. Lett. **89,** 156802 (2002).
46. V. Jovanov, T. Eissfeller, S. Kapfinger, E. C. Clark, F. Klotz, M. Bichler, J. G. Keizer, P. M. Koenraad, G. Abstreiter, and J. J. Finley, Phys. Rev. B **83,** 161303 (2011).
47. M. T. Björk, A. Fuhrer, A. E. Hansen, M. W. Larsson, L. E. Fröberg, and L. Samuelson, Phys. Rev. B **72,** 201307(R) (2005).
48. A. J. Bennett, M. A. Pooley, Y. Cao, N. Skold, I. Farrer, D. A. Ritchie, and A. J. Shields, Nat. Commun. **4,** 1522 (2013).
49. Y.-P. Shim, S. Oh, X. Hu, and M. Friesen, Phys. Rev. Lett. **106,** 180503 (2011).
50. F. Baruffa, P. Stano, and J. Fabian, Phys. Rev. Lett. **104,** 126401 (2010).
51. C. Lü, J. L. Cheng, and M. W. Wu, Phys. Rev. B **71,** 075308 (2005).
52. B. Shao, Y.-F. He, M. Feng, Y. Lu, and X. Zuo, J. Appl. Phys. **115,** 17A915 (2014).
53. D. Y. Li, Y. J. Zeng, L. M. C. Pereira, D. Batuk, J. Hadermann, Y. Z. Zhang, Z. Z. Ye, K. Temst, A. Vantomme, M. J. V. Bael, and C. V. Haesendonck, J. Appl. Phys. **114,** 033909 (2013).
54. J. H. Buß, J. Rudolph, F. Natali, F. Semond, and D. Hägele, Appl. Phys. Lett. **95,** 192107 (2009).
55. J. Lee, A. Venugopal, and V. Sih, Appl. Phys. Lett. **106,** 012403 (2015).
56. J. Zhang, R. Skomski, Y. F. Lu, and D. J. Sellmyer, Phys. Rev. B **75,** 214417 (2007).
57. K. V. Kavokin, Phys. Rev. B **64,** 075305 (2001).
58. K. V. Kavokin, Phys. Rev. B **69,** 075302 (2004).
59. L. P. Gor'kov and P. L. Krotkov, Phys. Rev. B **67,** 033203 (2003).
60. M. Venkatesan, C. B. Fitzgerald, J. G. Lunney, and J. M. D. Coey, Phys. Rev. Lett. **93,** 177206 (2004).
61. Rui Li and J. Q. You, Phys. Rev. B **90,** 035303 (2014).
62. S. Gangadharaiah, J. Sun, and O. A. Starykh, Phys. Rev. Lett. **100,** 156402 (2008).
63. I. Dzyaloshinskii, Phys. Chem. Solids **4,** 241 (1958).
64. T. Moriya, Phys. Rev. A **120,** 91 (1960).
65. E. I. Rashba, Sov. Phys. Solid State **2,** 1109 (1960).
66. M. I. Dyakonov and V. Y. Kachorovskii, Sov. Phys. Semicond. **20,** 110 (1986).
67. Y. Yamada, K. Ueno, T. Fukumura, H. T. Yuan, H. Shimotani, Y. Iwasa, L. Gu, S. Tsukimoto, Y. Ikuhara, and M. Kawasaki, Science **332,** 1065 (2011).

68. S. B. Ogale, Adv. Mater. **22,** 3125 (2010).
69. A. Jesche, R. W. McCallum, S. Thimmaiah, J. L. Jacobs, V. Taufour, A. Kreyssig, R. S. Houk, S. L. Budko, and P. C. Canfiel, Nat. Commun. **5,** 3333 (2014).
70. L.-Te Chang, C.-Yen Wang, J. Tang, T. Nie, W. Jiang, C.-Pu Chu, S. Arafin, L. He, M. Afsal, L.-J. Chen, and K. L. Wang, Nano Lett. **14,** 1823 (2014).
71. K. Yang, R. Wu, L. Shen, Y. P. Feng, Y. Dai, and B. Huang, Phys. Rev. B **81,** 125211 (2010).
72. P. Liu, S. Khmelevskyi, B. Kim, M. Marsman, D. Li, X.-Qiu Chen, D. D. Sarma, G. Kresse, and C. Franchini, Phys. Rev. B **92,** 054428 (2015).
73. G. Mattioli, F. Filippone, P. Alippi, and A. A. Bonapasta, Phys. Rev. B **78,** 241201 (2008).
74. H. Wu, A. Stroppa, S. Sakong, S. Picozzi, M. Scheffler, and P. Kratzer, Phys. Rev. Lett. **105,** 267203 (2010).
75. T. S. Herng, D.-C. Qi, T. Berlijn, J. B. Yi, K. S. Yang, Y. Dai, Y. P. Feng, I. Santoso, C. S.-Hanke, X. Y. Gao, A. T. S. Wee, W. Ku, J. Ding, and A. Rusydi, Phys. Rev. Lett. **105,** 207201 (2010).
76. X. Du, Q. Li, H. Su, and J. Yang, Phys. Rev. B **74,** 233201 (2006).
77. I. S. Elfimov, A. Rusydi, S. I. Csiszar, Z. Hu, H. H. Hsieh, H.-J. Lin, C.T. Chen, R. Liang, and G. A. Sawatzky, Phys. Rev. Lett. **98,** 137202 (2007).
78. K. Sato, L. Bergqvist, J. Kudrnovsky, P. H. Dederichs, O. Eriksson, I. Turek, B. Sanyal, G. Bouzerar, H. K.-Yoshida, V. A. Dinh, T. Fukushima, H. Kizaki, and R. Zeller, Rev. Mod. Phys. **82,** 1633 (2010).
79. K. Yang, Y. Dai, B. Huang, and Y. P. Feng, Phys. Rev. B **81,** 033202 (2010).
80. J. M. D. Coey, M. Venkatesan, and C. B. Fitzgerald, Nature **4,** 173 (2005).
81. H. Raebiger, S. Lany, and A. Zunger, Phys. Rev. Lett. **101,** 027203 (2008).
82. R. Janisch and N. A. Spaldin, Phys. Rev. B **73,** 035201 (2006).
83. P. Novak and F. R. Wagner, Phys. Rev. B **66,** 184434 (2002).
84. J. M. D. Coey, M. Venkatesan, P. Stamenov, C. B. Fitzgerald, and L. S. Dorneles, Phys. Rev. B **72,** 024450 (2005).

**Fig. 1.** (a) The structural dimensions, fields and spin wave propagation direction, the respective Cartesian coordinate axes, $\vec{M}_0$ and $\vec{H}_{ext}$ outside the shell, and an example vector plot of $\vec{M}_d(\vec{\rho})$ (the color map represents the absolute value of $\vec{M}_d(\vec{\rho})$, $\|\vec{M}_d(\vec{\rho})\|$), where a cylindrical core is perfectly cloaked from propagating magnons (outside the shell, $\vec{M}_d(\vec{\rho}) = sin(2\pi/(100nm) \times x)\hat{x} + cos(2\pi/(100nm) \times x)\hat{z}$) of $\vec{q} \perp \vec{M}_0$ type. $O$ represents the origin of the structure. The vector plot is magnified in the dashed box. (b) Schematic of magnetization precession in anisotropic $\bar{\bar{\gamma}}$. (c) Schematic of magnetization precession in anisotropic $\bar{\bar{M}}_s$. In (b) and (c), the distance between the red and blue curves indicate the amplitude of the gyromagnetic factor, while the black dashed curve indicates the trajectory of the magnetization $\vec{M}$. A thicker $\vec{M}$ indicates a higher saturation magnetization in (c).

**Fig. 2.** The magnetization in the $z$ direction ($M_z$) after 2.2 ns of an excitation with $\vec{h}_{mw} = 1 \times sin(2\pi/(50GHz) \times t)\hat{x}$ Oe at $x = 800$ nm. (a) No cylindrical core ($b = 0$ and $c = 0$). (b) Cylindrical core but no cloaking shell ($b = 50$ nm and $c = 50$ nm). (c) Cylindrical core with the shell designed for the $\bar{\bar{\gamma}}$-mechanism ($b = 100$ nm and $c = 50$ nm). (d) Cylindrical core with the shell designed for the $\bar{\bar{M}}_s$-mechanism ($b = 100$ nm and $c = 50$ nm). The dashed boxes represent the shadow region used for calculation of $M_{z,sr}(x)$ plotted in Fig. 3(a). The inner circle shows the boundary of the core. The larger circle shows the outer boundary of the shell.

**Fig. 3.** The variation of average $M_z$ ($M_{z,sr}(x) = \frac{1}{20nm} \int_{y_c-10nm}^{y_c+10nm} M_z(x,y)dy$) for (a) $0 \leq x \leq 420$ nm (the shadow region of the core), and (b) $0 \leq x \leq 820$ nm. The variation of average $M_z$ ($M_{z,T}(x) = \frac{1}{420nm} \int_{0nm}^{420nm} M_z(x,y)dy$) for (c) $0 \leq x \leq 420$ nm and (d) $0 \leq x \leq 820$ nm.

**Fig. 4.** Spatial pattern of (a) $\gamma_{rr}/\gamma_{\varphi\varphi}$, (b) $M_s(\vec{\rho})$ (color map) and $\vec{H}_{ext}(\vec{\rho})$ (cone plot), and (c) $\alpha_{zz}/\alpha_0$ for the $\bar{\gamma}$-mechanism. (d) The schematic of the ensemble of quantum dot molecules is in the left section. The schematic of an individual quantum dot molecule, its Cartesian coordination, the electric field ($E_{z,q}$), and magnetic field ($\vec{B}_q$) are in the right section. Spatial pattern of (e) $M_{s,rr}/M_{s,\varphi\varphi}$, and (f) $\alpha_{zz}/\alpha_0$ for the $\bar{M}_s$-mechanism. (g) Example schematic of a rutile crystalline structure consisting of metal and oxygen sites, as well as an interstitial impurity and oxygen vacancy positions. The right panel is an example of the octahedral oxygen coordination of an interstitial impurity. $\vec{E}_i$ is the external electric field on the supercell $i$ using Cartesian coordinates. Gray spheres represent metal (e.g. Sn, Hf, Ti etc.), red ones are O, green spheres are interstitial impurities (e.g. transition metals like V), and the hollow sphere is an oxygen vacancy $V_o$. (h) The schematic of the charge rings $\sigma_{1(2,3,4)}$ and the orbital moment directions $\vec{v}_{1(2,3,4)}$, for the four possible independent oxygen octahedral coordination of the interstitial impurities. (i) The schematic of the interaction of two adjacent interstitial impurity sites (*A* and *B*) with the nearby $V_o$, the resulting charge ring hybridization, and orbital moments. $\vec{v}_A$, $\vec{v}_B$, and $\vec{v}_T$ are the orbital moments of site *A*, site *B*, and the hybridization, respectively. It is assumed that $V_o$ is closer to the site *A* than to site *B*. In the right panel, the orbital moment vectors are shown with the same origin. In (a-c) and (e-f), the white regions in the center are hollow and no value is assigned to them.

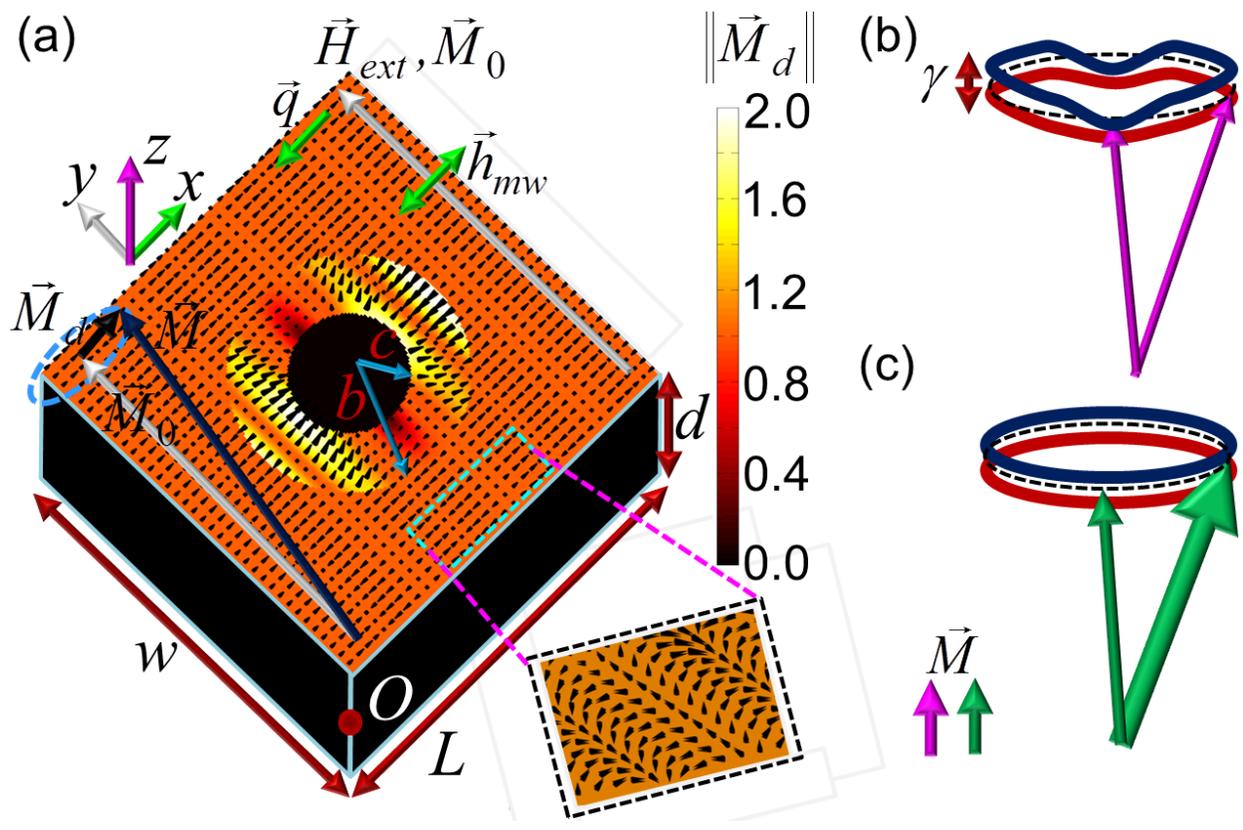

**Fig. 1**

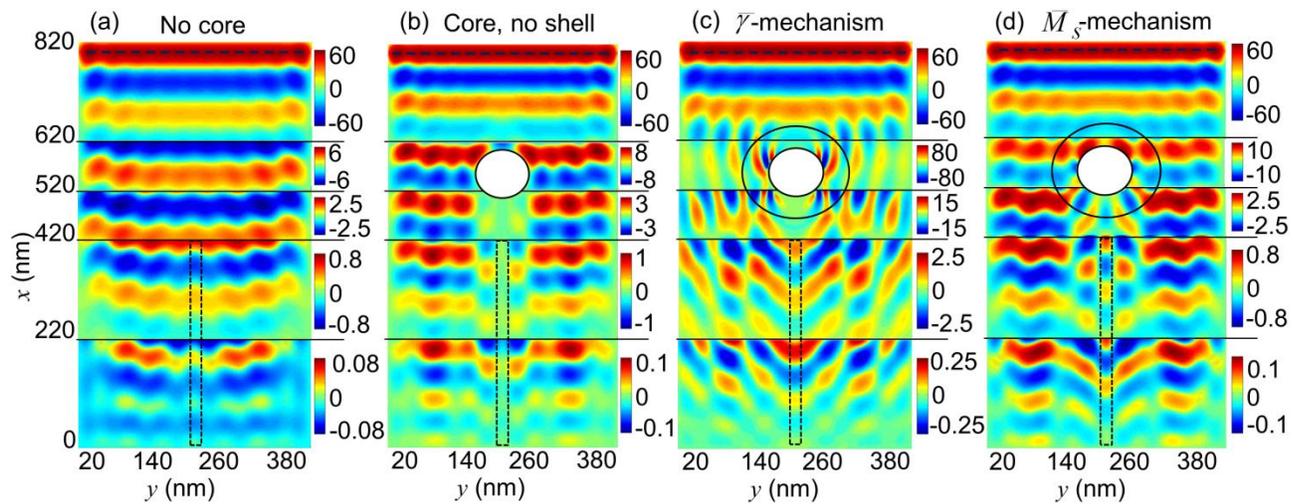

**Fig. 2**

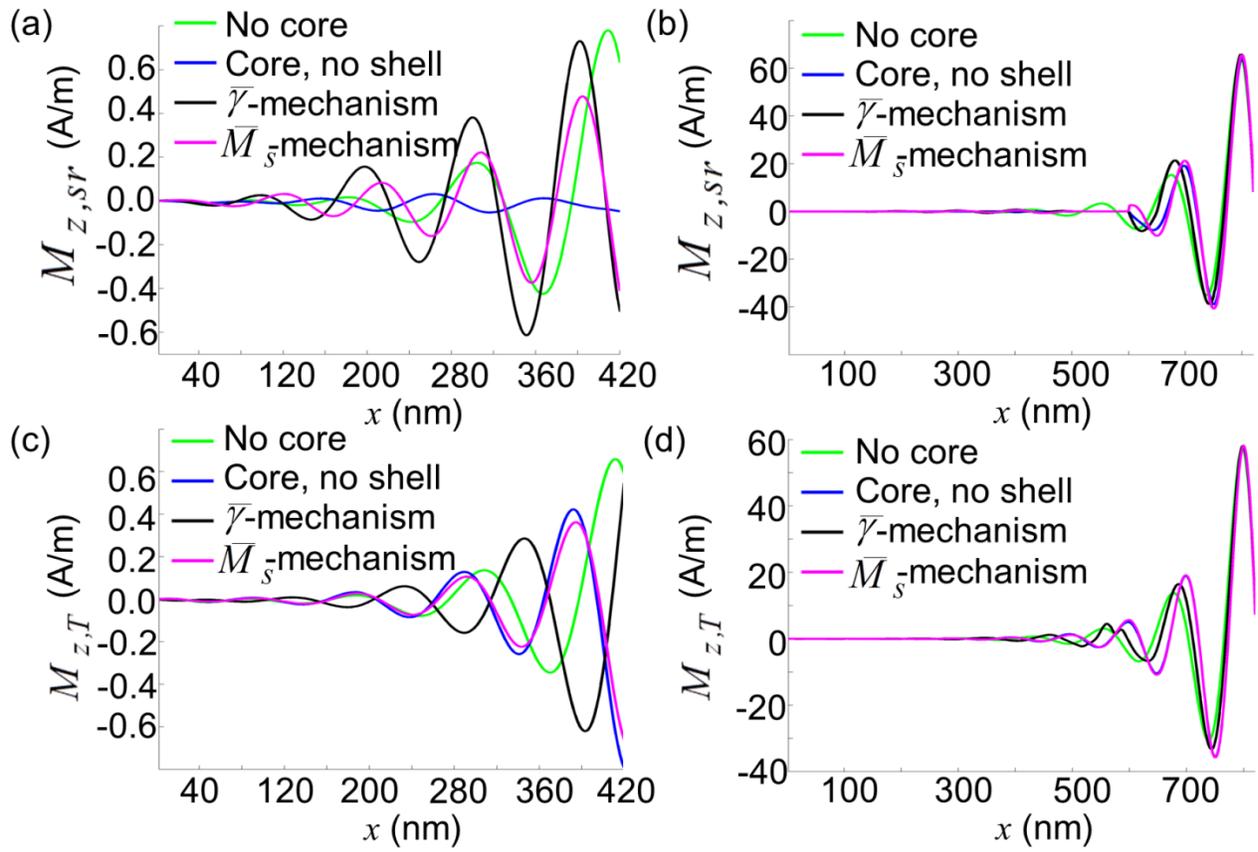

**Fig. 3**

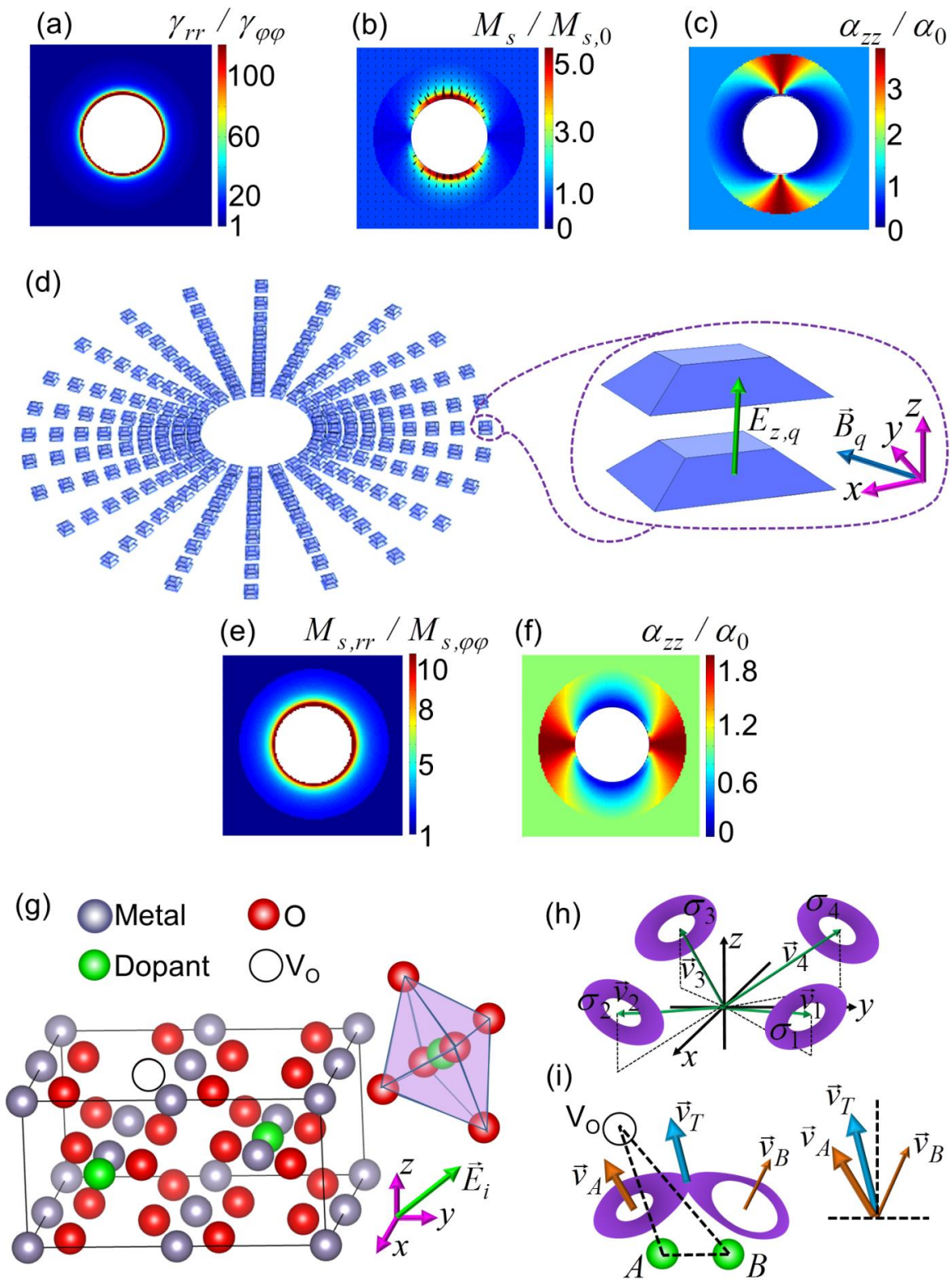

**Fig. 4**